\documentstyle[12pt]{article}
\def\be{\begin{equation}}
\def\ee{\end{equation}}
\def\bea{\begin{eqnarray}}
\def\eea{\end{eqnarray}}

\begin{document}

\begin{center}
{\Large{\bf Some Special Features of a Rotating-Moving D$p$-Brane }}

\vskip .5cm {\large Farzin Safarzadeh-Maleki } \vskip .1cm
{\sl f.safarzadeh@aut.ac.ir}\\
\end{center}

\begin{abstract}

Using the boundary string field theory (BSFT) techniques we study
the boundary state and partition function for a dynamical
(rotating-moving) D$p$-brane coupled to the electromagnetic and
tachyonic background fields in superstring theory. By making use
of the created partition function, the super BSFT action with a
tachyonic field and Dirac-Born-Infeld type action will be
constructed. By analyzing the obtained action interesting
features will be revealed.

\end{abstract}

{\it PACS numbers}: 11.25.-w; 11.25.Uv

{\it Keywords}: Rotating-Moving brane; Boundary state; Partition
function; BSFT; DBI.
 \vskip .5cm

\section{Introduction}

Many aspects of string/superstring theory have been discovered by
studying the essential objects of the theory name as D-branes
\cite {1,2}. By means of the boundary state method \cite
{3,4}, several properties and different configurations of
D-branes have been investigated \cite {5,6}. The
low-energy dynamics of a D-brane can be described by an effective
action, i.e., Dirac-Born-Infeld (DBI), plus its coupling to the
Ramond-Ramond fields. A DBI action is the unique invariant term
under the broken Lorentz symmetry in the lowest approximation. It
has the reparameterization invariance which removes the
longitudinal fluctuations of the brane and also is invariant
under the full space time symmetry \cite {7}.

On the other hand, in the context of background independent open
string field theory or BSFT for the case of superstring theory, an
effective spacetime action can be extracted by the corresponding
world sheet partition function. Moreover, these concepts have
been related to the boundary state formalism in an elegant way
\cite {8}-\cite {10}.

In this article by means of the BSFT techniques, we explore the boundary
state and partition function for a rotating-moving D$p$-brane coupled to
a $U(1)$ gauge potential in the world volume of the brane
and tachyonic background field in superstring theory.
Then by making use of the created partition function, we study the effective
action of such a system. In other words, we derive the super BSFT space time
action with a tachyonic field and DBI-type photonic field
associated with a rotating-moving brane from the world sheet
action.

The existence of the tachyonic boundary interaction breaks
conformal invariance and makes the theory off-shell. Therefore,
the BSFT would be a proper formalism to study such a dynamical
system. Being a relevant interaction, it will induce an RG flow
between two conformal field theories which describes tachyon
condensation \cite {9,11}.

The generalized BSFT action corresponding to this rotating-moving
D$p$-brane with background fields, enables us to gain a new
understanding of this dynamical system properties. Through
analyzing the action, among other interesting specifications such
as existence of a mixing structure between the tachyonic field
and rotating-moving term which affects the dynamics of the
action, by eliminating the effect of the zero mode portion of the
partition function an outstanding feature shows up: the rotation
does not generate any DBI-type factor, therefore, in spite of the
fact that we start from a rotating moving world sheet brane
action, we obtain the space time action without any contribution
of the angular momentum and the moment of inertia. This point will
be investigated both by means of the boundary state and partition
function approaches. Furthermore, by some simplifications, the
obtained action can be studied in two limits of small and large
values of tachyonic field. The former limit could generate the
right corrections of the effective action and the latter can
construct the conventional tachyon effective action describing
the dynamics of the tachyon field on a non-BPS Dp-brane of type
IIA or IIB superstring theory.

This article is organized as follows. In Sec. 2, the world sheet
action of a rotating-moving D$p$-brane with various background
fields in bosonic and superstring theory will be constructed. In
Sec. 3, The boundary state of such a dynamical will be extracted.
In Sec. 4, The partition function and DBI-type action associated
with a rotating-moving brane will be investigated.

\section{The actions}

In the context of BSFT or from the boundary sigma model point of
view for the closed string, in order to construct a dynamical
(rotating-moving) D-brane system in the presence of various
fields, we start with a bulk action \bea S_{\rm bulk} =
-\frac{1}{4\pi\alpha'} {\int}_\Sigma
d^{2}\sigma(\sqrt{-g}g^{ab}G_{\mu\nu}\partial_a X^{\mu}\partial_b
X^{\nu}), \eea and then couple the deformations to the original
theory via the boundary terms, as follows \bea S_{\rm
bdry}=\frac{1}{2\pi\alpha'} {\int}_{\partial\Sigma} d\sigma (
A_\alpha
\partial_{\sigma}X^{\alpha}+ \omega_{\alpha\beta}J^{\alpha\beta}_{\tau}
+iT^{2}(X^{\alpha})), \eea Therefore the desired action would be
$S^{\rm Bosonic}=S_{\rm bulk}+S_{\rm bdry}$, in which $\Sigma$ is
the closed string world-sheet, exchanged between the branes, and
$\partial\Sigma$ is its boundary which is a world sheet
space-like surface. This action contains a $U(1)$ gauge field
$A_\alpha$ which lives in the world-volume of the brane, an
$\omega$-term associated with the rotation and motion of the
brane and a tachyonic field. We shall use $\{X^\alpha|\alpha =0,
1, \cdot \cdot \cdot ,p \}$ for the world-volume directions of
the brane and $\{X^i| i= p+1, \cdot \cdot \cdot ,d-1\}$ for
directions perpendicular to it.

For simplifying the calculations the background field
$G_{\mu\nu}$ is considered to be constant. In addition, for the
$U(1)$ gauge field we apply the gauge
$A_{\alpha}=-\frac{1}{2}F_{\alpha \beta }X^{\beta}$ with constant
field strength. Besides, we use the following tachyon profile
$T^{2}=T_0 +\frac{1}{2}U_{\alpha\beta}X^{\alpha}X^{\beta}$, where
$T_0$ and the symmetric matrix  $U_{\alpha\beta}$ are constant.
Finally, the $\omega$-term, which is responsible for the brane's
rotation and motion, contains the anti-symmetric angular velocity
${\omega }_{\alpha \beta}$ and angular momentum density
$J^{\alpha \beta }_{\tau}$ which is given by ${\omega }_{\alpha
\beta}J^{\alpha \beta }_{\tau}=2{\omega }_{\alpha \beta
}X^{\alpha }{\partial }_{\tau }X^{\beta }$. In fact, the
component $\omega_{0 {\bar \alpha}}|_{{\bar \alpha} \neq 0}$
denotes the velocity of the brane along the direction $X^{\bar
\alpha}$ while $\omega_{{\bar \alpha}{\bar \beta}}$ represents
its rotation. Note that the rotation of the brane is considered
to be in its volume and its motion along the brane directions.
Moreover, according to the various fields inside the brane, the
Lorentz symmetry is broken and hence such a dynamic is meaningful.

The purpose of choosing the constant field strength and such a
tachyon profile is that for sufficiently small values, an
expansion of regularized action gives the right corrections and
one can study the influence of these couplings on the effective
action. On the other hand, because of losing the conformal
invariance, due to the presence of tachyon field, the influence
of these fields has been entered as corrections to the effective
action. The importance of this statement will be revealed in Sec.
4.

In order to supersymmetrize the above action one should add
fermionic partners of the $X^{ \mu }$'s , i.e., ${\psi }^{\mu
}$'s such that the supersymmetrized action would be invariant
under the global worldsheet supersymmetry. Steps of
supersymmetrizing the bosonic action using a superspace formalism
can be summarized as: first, using the super world-sheet
coordinates and hence, adding the Grassmanian integration
measurement; second, changing the ordinary derivative to the
super covariant derivative and third, changing the fields to the
superfields which are the quantum mechanical degrees of freedom
living on the boundary and the restrictions to the boundary of the
standard world sheet super coordinate \cite {1, 8, 12}.
Therefore, the supersymmetrized version of the rotating-moving
world sheet action with photonic and tachyonic background fields
would be $ S_{\rm bdry+\rm bulk}^{\rm Super}= S^{\rm
Fermionic}+S^{\rm Bosonic}$, in which \bea S^{\rm Fermionic}&=&
\frac{-i}{4\pi\alpha'} {\int}_{\partial\Sigma} d\sigma \left[
\theta^{\alpha} F_{\alpha \beta} \theta^{\beta}+
4\omega_{\alpha\beta}[( \psi ^{\alpha}_{+}+i\eta
\psi^{\alpha}_{-})( \psi ^{\beta}_{+}-i\eta
\psi^{\beta}_{-})]+i(\theta^{\alpha}\partial_{\alpha}T)\partial_{\sigma}^{-1}
(\theta^{\beta}\partial_{\beta}T)\right]\
\nonumber\\
&+& \frac{i}{4\pi\alpha'} {\int}_\Sigma
d^{2}\sigma(\bar{\psi}^{\mu} \rho^{a}\partial_a \psi^{\mu}), \eea
where the boundary fermion $\theta^{\alpha}=\psi
^{\alpha}_{+}+i\eta \psi^{\alpha}_{-} $ is the linear combination
of the solution of the closed superstring equations of motion $
\psi^{\alpha}= \left(
\begin{array}{cc}
\psi^{\alpha}_{+} \\
\psi^{\alpha}_{-}\end{array} \right)$. Moreover,
$\partial_{\sigma}^{-1}$ is a Green function that could be
written in terms of the sign function
$\partial_{\sigma}^{-1}g(\sigma)=\frac{1}{2}\int {d\sigma' \rm sgn
(\sigma-\sigma')g(\sigma')}$ and $\rho^{a}$'s are two dimensional
Dirac matrices. We use the same notation as \cite {10}.

\section{Boundary state of the system}

By means of the obtained action, one can define the associated
bosonic boundary state as \bea &~& |B;S^{\rm bdry}_{\rm
Bos}\rangle \ = \int{[dx\ d\overline{x}}]\ e^{iS^{\rm bdry}_{\rm
Bos}[ x,\overline{x}]}|x,\overline{x}\rangle
\nonumber\\
&~& |x,\overline{x}\rangle =\prod^{\infty }_{m=1} {{\exp
(-\frac{1}{2}\ }}{\overline{x}}_mx_m- a^{\dagger
}_m{\tilde{a}}^{\dagger }_m +a^{\dagger
}_mx_m+{\overline{x}}_m{\tilde{a}}^{\dagger }_m)\ |vac\rangle,
\eea where $x_m=a_m e^{-2im\tau}+\tilde{a}^{\dagger }_m
e^{2im\tau}$, defined in terms of the bosonic oscillators, can be
derived from the solution of closed superstring equations of
motion \bea X^{\mu }\left(\sigma ,\tau \right)=x^{\mu
}_0+2{\alpha }^{{\rm '}}p^{\mu }\tau +\ \sqrt{\frac{{\alpha
}^{{\rm '}}}{2}}\sum_{m>0}{m^{-\frac{1}{2}}}(x^{\mu
}_me^{2im\sigma }+{\overline{x}}^{\mu }_me^{-2im\sigma }).\eea
Beside, the bosonic boundary action $\ S^{\rm bdry}_{\rm
Bos}=(S^{\rm F}+S^{\rm T}+S^{\rm \omega })_{\rm Bos}$ is related
to the photon, tachyon and the rotation terms as follows \bea &~&
S^{\rm F}_{\rm Bos}=\frac{i}{2} \sum_{m>0}{F_{\alpha \beta \
}{\overline{x}^{\alpha }_m}x^{\beta }_m},
\\
&~& S^{\rm T}_{\rm Bos}=\frac{i}{4{\alpha }^{{\rm '}}}U_{\alpha
\beta }x^{\alpha }_0x^{\beta }_0+\frac{i}{4}
\sum_{m>0}{\frac{U_{\alpha \beta \ }}{m}{\overline{x}}^{\alpha
}_m x^{\beta }_m} ,\\
&~& S^{\rm \omega }_{\rm Bos}=2{\omega }_{\alpha \beta \
}x^{\alpha }_0p^{\beta }-2i{\omega }_{\alpha \beta \
}\sum_{m>0}{({\overline{x}}^{\alpha }_m{a}^{\beta }_m+x^{\alpha
}_m  \tilde{a}^{ \beta }_m )}. \eea

The above equations are obtained by applying the oscillating modes
in the bosonic sector. Thus by inserting the obtained $\ S^{\rm
bdry}_{\rm Bos}$ in Eq. (4) the solution of the bosonic part
would be extracted as \bea |B;S^{\rm bdry}_{\rm Bos}\rangle
&=&\frac{T_p}{2}\prod^{\infty }_{n=1} {[\det
Q_{(n)}]^{-1}}\;{\exp \left[-\sum^{\infty }_{m=1}
{\frac{1}{m}{\alpha }^{\mu }_{- m}S_{(m)\mu \nu }
{\widetilde{\alpha }}^{\nu }_{-m}}\right]\ } {|0\rangle}_{\alpha}
\otimes {|0\rangle}_{\widetilde{\alpha }} \;
\nonumber\\
&\times& \int^{\infty }_{{\rm -}\infty } \exp\left\{{\alpha
}^{{\rm '}}\left[\sum^{p}_{\alpha  =0} {\left(U^{{\rm -}{\rm
1}}{\mathbf A}\right)}_{\alpha \alpha}
{\left(p^{\alpha}\right)}^{{\rm 2}}{\rm +} \sum^{p}_{\alpha
,\beta {\rm =0},\alpha \ne \beta}{{\left(U^{{\rm -}{\rm
1}}{\mathbf A}+{\mathbf A}^T U^{-1}\right)}_{\alpha \beta }
p^{\alpha }p^{\beta}}\right]\right\}{\rm \ \ }
\nonumber\\
&\times& \left( \prod_{\alpha}{\rm |}p^{\alpha}\rangle
dp^{\alpha}\right) \otimes\prod_i{\delta {\rm (}x^i}{\rm
-}y^i{\rm )} {\rm |}p^i{\rm =0}\rangle . \eea in which the
matrices are
 \bea &~& Q_{(n){\alpha \beta }} =
{\eta }_{\alpha \beta }-F_{{\mathbf \alpha }{\mathbf \beta
}}-\frac{1}{2n}U_{\alpha \beta },
\nonumber\\
&~& S_{(m)\mu\nu}=(\Delta_{(m)\alpha \beta}\; ,\; -{\delta}_{ij}),
\nonumber\\
&~& \Delta_{(m)\alpha \beta} = (M_{(m)}^{-1}N_{(m)})_{\alpha
\beta},
\nonumber\\
&~& M_{(m){\alpha \beta }} = {\eta }_{\alpha \beta }+4{\omega
}_{\alpha \beta }-F_{{\mathbf \alpha }{\mathbf \beta
}}-\frac{1}{2m}U_{\alpha \beta },
\nonumber\\
&~& N_{(m){\alpha \beta }} = {\eta }_{\alpha \beta } +4{\omega
}_{\alpha \beta } +F_{{\mathbf \alpha }{\mathbf \beta }}
+\frac{1}{2m}U_{\alpha \beta },
\nonumber\\
&~& \mathbf A}_{\alpha \beta}=\eta_{\alpha \beta} +
4\omega_{\alpha \beta . \eea

Similarly for the fermionic boundary state we have \bea
&~&|B;S^{\rm bdry}_{\rm Ferm}\rangle \ = \int{[d\theta\
d\overline{\theta}}]\ e^{iS^{\rm bdry}_{\rm Ferm}[
\theta,\overline{\theta}]}|\theta,\overline{\theta}\rangle
\nonumber\\
&~&|\theta,\overline{\theta}\rangle =\prod^{\infty }_{m=1} {{\exp
(-\frac{1}{2}\ }}{\overline{\theta}}_m\theta_m+i\eta \psi^{\dagger
}_m{\tilde{\psi}}^{\dagger }_m +\psi^{\dagger
}_m\theta_m-i\eta{\overline{\theta}}_m{\tilde{\psi}}^{\dagger
}_m)\ |vac\rangle, \eea where $\theta_m=\tilde{\psi}
_{m}e^{-2im\tau}-i\eta \psi^{\dagger }_{m}e^{2im\tau}$, described
as a combination of fermionic oscillators, could be found as \bea
\theta^{\mu}=\sum_{m>0}
(\theta^{\mu}_me^{-2im\sigma}+\bar{\theta}^{\mu}_me^{2im\sigma}).
\nonumber \eea

The fermionic boundary action $\ S^{\rm bdry}_{\rm
Ferm}=(S^F+S^T+S^{\omega })_{\rm Ferm}$ contains \bea &~& S^{\rm
F}_{\rm Ferm}=\frac{i}{2{\alpha }^{{\rm '}}} {F_{\alpha \beta \
}{\overline{\theta}}^{\alpha }_0}\theta^{\beta
}_0+\frac{i}{2{\alpha }^{{\rm '}}} \sum_{r>0}F_{\alpha \beta \
}{\overline{\theta}}^{\alpha
}_r\theta^{\beta }_r,\\
&~& S^{\rm T}_{\rm Ferm}=\frac{i}{4} \sum_{r>0}{\frac{U_{\alpha
\beta }}{r}{\overline{\theta}}^{\alpha }_r\theta^{\beta }_r}, \\
&~& S^{\rm \omega }_{\rm Ferm}=2i{\omega }_{\alpha \beta \
}\sum_{r>0}{({\overline{\theta}}^{\alpha }_r{\tilde{\psi}}^{\beta
}_r+i\eta \psi^{ \beta }_r \theta^{\alpha }_r)}. \eea

Therefore, the NS-NS and R-R sectors possess the following
fermionic boundary states \bea &~& |B^{ \rm bdry}_{\rm Ferm}
\rangle_{\rm NS}=\prod^{\infty}_{r=1/2}[\det Q_{(r)}]\exp
\bigg{[}i\sum^{\infty}_{r=1/2}(b^{\mu }_{-r} S_{(r)\mu
\nu}{\widetilde b}^{\nu}_{-r})\bigg{]}|0
\rangle ,\\
&~& |B^{ \rm bdry}_{\rm Ferm} \;\rangle_{\rm R} =\prod^{\infty
}_{n=1}[\det Q_{(n)}] {\exp \left[i \sum^{\infty }_{m=1}{(d^{\mu
}_{-m}S_{(m)\mu \nu } {\widetilde{d}}^{\nu }_{-m})} \right]\ }
|B\rangle^{(0)}_{\rm R}. \eea

The role of the explicit form of zero-mode state
$|B\rangle^{(0)}_{\rm R}$ is not important here. This is because
for obtaining the partition function, this state would be
projected onto the bra-vacuum. The interested reader could find
its explicit form in \cite {6}.

It should be noted that in above calculations ${\alpha }^{{\rm
'}}=1$ has been considered.
\section{ Partition function and DBI-type action}

It has been demonstrated that the normalization factor of the
boundary state is the well-known, DBI Lagrangian. On the other
hand, by means of the partition function one can obtain the
DBI-type action. This delicate relation can be described as: at
the tree level the disk partition function in the BSFT appears as
the normalization factor of the boundary state. The partition
function can be obtained by the vacuum amplitude of the boundary
state \bea Z^{disk}=\langle vacuum|B\rangle. \eea

Thus, in our setup the partition function possesses the following
feature \bea Z_{\rm Bos}^{\rm Disk} &=& \frac{T_p}{2}\int^{\infty
}_{{\rm -}\infty } {\prod_{\alpha }{dp^{\alpha
}}}\exp\bigg{\{}{{\alpha }^{{\rm '}}\left[\sum^{p}_{\alpha =0}
{\left(U^{{\rm -}{\rm 1}}{\mathbf A}\right)}_{\alpha \alpha}
{\left(p^{\alpha}\right)}^{{\rm 2}}{\rm +} \sum^{p}_{\alpha
,\beta {\rm =0},\alpha \ne \beta}{{\left(U^{{\rm -}{\rm
1}}{\mathbf A}+{\mathbf A}^T U^{-1}\right)}_{\alpha \beta }
p^{\alpha }p^{\beta}}\right]\bigg{\}}{\rm \ \ }}
\nonumber\\
& \times & \prod^{\infty }_{n=1}{[\det Q_{(n)}]^{-1}}. \eea for
the bosonic part of the partition function, and \bea Z_{\rm
Ferm}^{\rm Disk}=\prod^{\infty}_{k>0}[\det Q_{(k)}], \eea for the
fermionic part, where $k$ is half-integer (integer) for the NS-NS
(R-R) sector. Considering both fermionic and bosonic parts, the
total partition function in superstring theory is given by \bea
Z_{\rm total}^{\rm Disk} &=& \frac{T_p}{2}Z_{\rm (0)}^{\rm
Disk}\frac{\prod^{\infty}_{k>0}[\det Q_{(k)}]}{\prod^{\infty
}_{n=1} {[\det Q_{(n)}]}}, \eea Where \bea Z_{\rm (0)}^{\rm
Disk}\equiv\int^{\infty }_{{\rm -}\infty } {\prod_{\alpha
}{dp^{\alpha }}}\exp\bigg{\{}{{\alpha }^{{\rm
'}}\left[\sum^{p}_{\alpha =0} {\left(U^{{\rm -}{\rm 1}}{\mathbf
A}\right)}_{\alpha \alpha} {\left(p^{\alpha}\right)}^{{\rm
2}}{\rm +} \sum^{p}_{\alpha ,\beta {\rm =0},\alpha \ne
\beta}{{\left(U^{{\rm -}{\rm 1}}{\mathbf A}+{\mathbf A}^T
U^{-1}\right)}_{\alpha \beta } p^{\alpha
}p^{\beta}}\right]\bigg{\}}{\rm \ \ }}\eea

In order to prevent a divergent infinite product, let us use the
$\zeta$-function regularization \bea Z_{\rm total}^{\rm Disk}
\longrightarrow \frac{T_p}{2}Z_{\rm (0)}^{\rm Disk}\sqrt{
\det(\eta -F)}\det \left[\frac{\sqrt{\pi}\Gamma \left(
1-\frac{1}{2}(\eta -F)^{-1}U \right)}{\Gamma
\left(\frac{1}{2}-\frac{1}{2}(\eta -F)^{-1}U \right) }\right].
\eea where the arrow sign instead of equality has been used to
show the application of this useful function.

The structure of the effective action can be formed by the
interaction/boundary terms that are coupled to the world sheet
action. The tree level closed string effective action can be
studied by string partition on the sphere and in view of open
string theory this action is described by the disk partition
function. In superstring theory, the leading term in the latter
case is called the Born-Infeld action \cite {12}. Therefore, the
extension of the DBI action or in other word, the super BSFT
action with a tachyonic field and DBI-type photonic field
associated with this dynamical brane would appear as \bea
S=\frac{T_p}{2}(\pi) ^{(p+1)/2}\int{d^{p+1}\xi \frac{\sqrt{
\det(\eta -F)}}{\sqrt{{\det (D + H)}}} \det
\left[\frac{\sqrt{\pi}\Gamma \left( 1-\frac{1}{2}(\eta -F)^{-1}U
\right)}{\Gamma \left(\frac{1}{2}-\frac{1}{2}(\eta -F)^{-1}U
\right) }\right]},\eea in which we have used the zero mode
partition function after integrating on the momenta, where the
diagonal matrix possesses the elements $D_{\alpha \beta}=
(U^{-1}{\mathbf A})_{\alpha \alpha}\delta_{\alpha \beta}$, the
the matrix $H_{\alpha \beta}$ is defined by \bea H_{\alpha
\beta}= \bigg{\{}
\begin{array}{c}
(U^{-1}{\mathbf A} + {\mathbf A}^T U^{-1})_{\alpha \beta}\;\;,\;\;\;\alpha \neq \beta ,\\
0 \;\;\;\;\;\;\;\;\;\;\;\;\;\;\;\;\;\;\;\;\;\;\;\;\;\;\;\;\;\;
\;,\;\;\;\alpha = \beta .
\end{array}
\eea and ${\mathbf A}$ has been introduced in Eq. (10). In
addition, due to the tachyon profile we have $(\eta
-F)^{-1}U=2[(\eta
-F)^{-1}]^{\alpha\beta}\partial_{\alpha}T\partial_{\beta}T$, note
that $T_0=0$ is considered for simplicity.

According to the above relations, the rotating-moving effect of
the described brane can be seen in the zero mode part of the
effective action (23). This is due to the fact that according to
BSFT formulation for constructing a DBI Lagrangian, the
oscillating part of the boundary state which contains the
dynamical term, does not generate any DBI factor and hence the
rotating-moving dynamic of this brane action lies in its zero
mode contribution.

An interesting feature of this set up is that by applying the
boundary state procedures of calculations, explicitly one can
obtain that the tachyon and rotating-moving terms have a mixed
structure. Using the eigenvalues obtained from the zero mode
boundary state equation, one can deduce $p^\alpha =-\frac{1}{2}[
( {\mathbf A} )^{-1}U ]^\alpha_{\;\;\beta} x^\beta$, which means
along the world volume of the brane, momentum of the closed
string depends on its center of mass position. Consequently, in
the presence of the tachyon field the emitted closed string feels
an exotic potential which affects its evolution. Now consider
such a system without the tachyon, then by calculating the
boundary state equation of the zero mode part one would obtain
$2{\mathbf A}_{\alpha \beta }p^{\beta }{|B\rangle}^{(0)} =0$,
which shows that this case does not have a mixed structure and
hence its zero mode dynamic would be independent of the
background field. Therefore unlike the one with tachyonic field,
with technical difficulties in calculations, a much more simple
case would be appeared.

According to the non-BPS characteristic of this dynamical D-brane
and its special background fields, in order to insert its
contribution from the fermionic fields, one should consider its
property as a $(1, 0)_p$-system. Note that a generic $(n,m)$
could define as $(n,m)_p=n|Bp \rangle_{\rm NS}+m|Bp \rangle_{\rm
R}$, for combinations of GSO projected states on behalf of both
the Neumann boundary conditions in $p+1$ and Dirichlet boundary
conditions for $d-p-1$ dimensions. Thus the fermionic
contribution is that of the NS-NS sector.

It has been demonstrated that the tachyon kinks, tubes, and
vortices could be found by studying the equations of motion from
the DBI-type effective field theory \cite {14}. In other word,
inhomogenous solutions of the equations of motion which follow
from the action encode non-trivial information about the decay of
higher dimensional branes into lower dimensional ones. Therefore,
one can check the tachyon kinks as solutions of classical
equations of motion of this rotating-moving system. Specifically,
for the case in hand the spectrum of the kinks solutions becomes
richer according to our special background fields.

Now let us simplify the obtained action and drop the effect of
zero mode partition function $Z_{\rm (0)}^{\rm Disk}$.
Consequently we have \bea S=\frac{T_p}{2}\int{d^{p+1}\xi \ V({\rm
Tachyonic})\sqrt{ \det(\eta -F)} \det
\left[\frac{\sqrt{\pi}\Gamma \left( 1-\frac{1}{2}(\eta -F)^{-1}U
\right)}{\Gamma \left(\frac{1}{2}-\frac{1}{2}(\eta -F)^{-1}U
\right) }\right]}.\eea

Some properties of the above action are as follow:

-The kinetic part of the action is,
$\det\left[\frac{\sqrt{\pi}\Gamma \left( 1-\frac{1}{2}(\eta
-F)^{-1}U \right)}{\Gamma \left(\frac{1}{2}-\frac{1}{2}(\eta
-F)^{-1}U \right) }\right]$ and the potential term is tacitly
proportional to $V({\rm Tachyonic})= e^{-\frac{i}{2}T^2}$ due to
the defined tachyon profile. It should be noted that the exact
tachyon potential is obtained by setting $U_{\mu} = 0$ in
tachyonic profile and computing the path integral boundary action
\cite {8, 13, 14}.

- This action is a generalized form of a non-BPS D9-brane action
in \cite {8, 12} which is an interpolating action that coincide
with the low-energy two-derivative effective action for a
D25-brane in open bosonic string theory and the corresponding
expression for the non-BPS D9-brane in type IIA theory, in the
two-derivative approximation.

- In the beginning we start from a rotating-moving world sheet
brane action but at the end we reach to the space time action
without any contribution of the angular momentum and the moment
of inertia. In other word, why does dynamic of this typical brane
effective action lie in its zero mode portion? In the following
two interpretations of this case are explained: The bosonic
boundary state (9) consists of oscillating and zero mode parts.
The first line in (9) with the infinite determinant and the
exponential factor is the contribution of the oscillators which
act on the vacuum of oscillators. The remaining part belongs to
the zero modes, a delta function which has been included to fix
the location of the D-brane by imposing an extra condition on the
position operator in the transverse direction and a momentum
dependent integral which comes from taking the zero mode action
into account. As it seems, there is no $\delta$-function in the
directions in which we perform the rotation and therefore the
rotation acts on the zero modes. As a result, by eliminating the
effect of the zero mode portion, the rotation does not generate
any DBI-type factor (boundary state approach). This case is
similar to that of the rotation in a delocalized Dp-brane in
\cite {5}. In other point of view, this is because of the
anti-symmetric characteristic of the rotating-moving
($\omega$)-term. As a matter of fact, this portion would be
eliminated in the path integral calculations and therefore it
does not have any contribution in this action (partition function
approach).

- The obtained action can be checked in the limits of small and
large values of tachyon field. Where the first limit could
reproduce the right corrections of the effective action due to
the special boundary deformation which caused the conformal
invariance to be lost, and the latter that can construct the
conventional tachyon effective action describing the dynamics of
the tachyon field on a non-BPS D$p$-brane of type IIA or IIB
superstring theory. In the following these statements would be
clarified: For the first case, in the kinetic term of action (25)
consider $R(-\frac{1}{2}(\eta-F)^{-1}U)= \left[\frac{\sqrt{\pi}
\Gamma \left( 1-\frac{1}{2}(\eta-F)^{-1}U \right)} {\Gamma
\left(\frac{1}{2}-\frac{1}{2}(\eta-F)^{-1}U \right) }\right]$.
Now by defining \bea z\equiv-\frac{1}{2}(\eta -F)^{-1}U,\eea
$R(z)$ could be rewritten as \bea
R(z)=\frac{4^{z}z\Gamma{(z)}^2}{2\Gamma{(2z)}}=2^{2z-1}z B(z),\eea
in which $B(z)$ is the Beta function. One can reproduce the
leading term by expanding $R(z)$ around $z=0$, hence the above
action takes the following form \bea
S=\frac{T_p}{2}\int{d^{p+1}\xi \ V({\rm Tachyonic})
\sqrt{\det(\eta -F)} \det\left[1 + Log[4]z + (-\frac{\pi^2}{6} +
2Log[2]^2)z^2 + O[z]^3\right]}\eea It can be interpreted as the
corrections to the DBI-type action according to the special
format of the kinematic part of the D$p$-brane Lagrangian.

In order to construct the tachyonic effective action of a non-BPS
D$p$-brane of type IIA or IIB superstring theory, let us check
action (25) for large values of $z$. Thus, by means of the
recursion relation obtained from $\Gamma(1+y)=y\Gamma(y)$ for the
Beta function and applying the Stirling approximation $B(y)\simeq
2^{1-2y}\sqrt{\frac{\pi}{y}}$ for large $y$, one can derive \bea
R(z)\simeq
\frac{(z+\frac{2s-1}{2})(z+\frac{2s-3}{2})...(z+\frac{1}{2})}
{(z+s-1)(z+s-2)...(z+1)} \sqrt{\frac{\pi}{z+s}};\;\;\;\;\;
s=1,2,3,....\eea Now in order to reach an action with a familiar
structure let \bea R(z)\mid_{s=1}\simeq\sqrt{\pi}\sqrt{1+z},\eea
for $1+z\gg1$. Returning Eq. (30) into the effective action (25),
applying the determinant expansion $\det{(1+y)}\approx1+{\rm
Tr}y$ and omitting the total field strength $F$ for simplicity,
an intimate result will be revealed \bea S\approx
\frac{T_p}{2}\sqrt{\pi}\int{d^{p+1}\xi \ V(T) \sqrt{1-
\eta^{\alpha\beta}\partial_{\alpha}T\partial_{\beta}T}},\eea This
action is similar to the proposed tachyon effective action for
large $T$ on an unstable D$p$-brane system in \cite {13}.


\end{document}